# Hox genes underlie metazoan development, but what controls them?

**Raffaele Di Giacomo[1°], Bruno Maresca[2°] and Jeffrey H. Schwartz[3*]**

[1]Department of Mechanical and Process engineering 5 (D-MAVT), Swiss Federal Institute of Technology (ETH), Zurich, CH, Switzerland

[2]Dept. of Pharmacy, Division of BioMedicine, University of Salerno, Italy.

[3]Departments of Anthropology and History and Philosophy of Science, University of Pittsburgh, USA.

[°]These authors contributed equally to this work.   [*]Corresponding author: Email: jhs@pitt.edu

**Abstract**  Although metazoan development is conceived as resulting from gene regulatory networks (GRNs) controlled by Hox genes, a better analogy is computer architecture: i.e., a task accomplished in sequential steps linked to an external referent that "counts" each step. A developmental "step" equals the expression of genes in specific cells at specific times and telomeres represent external "counters" wherein "counting" is a function of telomere shortening at each cell division that permits the sequential expression of Hox genes and, ultimately, complex form. Metazoan development thus best resembles a Turing machine, which could be used to model the development of any metazoan.

**Introduction** In metazoan development complexity emerges regardless of taxic differences in cell number and growth rates of different tissues and organs. To explain this, Wolpert (Wolpert 1969) suggested that specific cells secrete a graded concentration of morphogens such that, de-pending on the distance of target genes from a morphogen source, their differential expression above and below certain thresholds generates a particular spatial patterning of cell differentiation. Support of this model was sought in *Bicoid*, which is a homeodomain-containing transcription factor (TF) that in *Drosophila* embryos determines the anteroposterior axis (Wartlick, et al. 2009), and controls activation of the segmentation gene hunchback anteriorly (Okabe-Oho, et al. 2009; Schultz and Tautz 1995). By transmitting information across cells, morphogens were seen as orchestrating the patterning of organismal structure by forming long range gradients in which cells "learn" their relative spatial positions. Recently, Wolpert (Wolpert 2011) rejected this model. Since morphogen diffusion depends on a tissue's three-dimensional structure, including density and nature of cell-cell contacts, the binding of diffusible molecules to receptors makes problematic determining effective morphogen concentrations. Moreover, since the properties of morphogen gradients are affected by other molecules and change over time (Yan and Lin 43 2009), pattern formation is affected pleiotropically (Wolpert 2007). Consequently, Wolpert (Wolpert 2011) (p. 364) concluded that "extracellular diffusion is not reliable enough to specify positional information, and other mechanisms must be involved." Instead, he (Wolpert 2010) proposed that 1) stem cells may possess a mechanism that counts successive cell divisions and 2) telomeres may be involved. We (Di Giacomo, et al. 2013) hypothesized that the emergence of complex, metazoan structure was the result of first increasing telomere length, and then shifting to a telomerase-negative cellular state in which telomeres shorten with each cell division, thereby creating a linearly changing reference point (i.e. an irreversible counting mechanism) that differentially exposes previously silenced Hox genes. The basic metazoan "Bauplan" is, therefore, the consequence of a sequential, stepwise, time-and spacedependent, expression of Hox genes inextricably linked to telomere shortening. Here we argue, since the metazoan

Bauplan is the consequence of a series of sequential steps performing a specific task with reference to an external point "counting" each step, that 1) metazoan development is similar to computer architecture, and 2) the existence of an external counter implies that it can be conceived as either a universal or a series of special Turing machines that can be used to model the development of any metazoan.

**Telomeres** In contrast to the linear chromosomes of eukaryote cells, the single, circular chromosome of prokaryote cells permits complete replication of its DNA with every cell division. For some number of early post-zygotic mitotic divisions in lower and higher eukaryotic cells, initial chromosome length is maintained because cells express the enzyme telomerase, which adds six base-pair repeat sequences to the chromosomal ends. In higher eukaryotic cells, chromosomes then begin to shorten because they down-regulate telomerase, becoming telomerase negative (O'Hare and Delany 2005; Taylor and Delany 2000). In turn, the intrinsic mechanism of DNA polymerase prevents complete replication of the lagging strand, resulting in shortening of its 3' end. Consequently, with each cell division, one daughter cell retains a complete copy of the parental cell's chromosome while the other receives a telomerically shortened chromosome, such that the chromosomes of descendants of one cell lineage remain long, while those of descendants of the other cell lineage continually shorten. Cell division thus generates a branching pattern that reflects the distribution of telomeres of varying length in different cells (fig. 1 above). Further, as observed in four normal human diploid cells (Huffman, et al. 2000), the rate of telomere shortening is proportional to the 3' overhangs, which are of differing lengths in different adult tissues.

**Hox gene expression** Metazoan development requires expression of a finely tuned genetic program. Hox genes code for TFs that govern pattern formation along anteroposterior and bilateral embryonic axes (Mallo, et al. 2010) occur in clusters toward one of the chromosomal telomeric regions, and their relative positions determine their differential expression (Duboule 1994). In embryos of genetically modified mice, when Hox genes normally expressed later in development are relocated nearer the telomeric extremity, they are expressed earlier than those closer to the centromere (Tschopp and Duboule 2011). In humans, alteration of (partially identified) repetitive sequences in subterlomeric regions results in various teratological conditions (Heutink, et al. 1994). Further, changes in the configuration of a Hox gene cluster affect the transcriptional regulation of neighboring genes, e.g. a mutation in the human HoxD13 results in digital malformation (synpolydactyly) (Quinonez and Innis 2014). Consequently, the fundamentals of development common to all metazoans – Gene Regulatory Networks (GRNs) (Davidson and Erwin 2006) – are the consequence of a time-dependent expression of Hox genes, which must be properly ordered and in the correct physical location. *Saccharomyces cerevisiae* (yeast) telomeres play a major role in determining reversible mating type (Lebrun, et al. 2001): i.e. *MATa* and *MATα* are TFs that induce specific mating type genes. In the diploid state, yeast telomerase is expressed and Sir1-Sir4 proteins fold telomeres and silence transcription at the mating-type locus [telomeric position effect (TPE)]. In the haploid state, yeast telomerase is not expressed and the absence of Sir2 prevents telomere folding, thereby enabling gene conversion (= mating-type switching) from the silent locus near the telomeric end to an alternate gene type in a transcriptionally active locus closer to the centromere. In humans, SIRT6 – the equivalent of yeast Sir2 (Imai, et al. 2000) – is a histone deacetylase that regulates chromatin structure and telomere functionality through deacetylation of a specific histone tail residue (H3K9) that promotes formation of specialized telomeric chromatin

(Michishita, et al. 2009). As demonstrated in *S. cerevisiae*, master-and proto-silencers in the subtelomeric region reinforce TPE (Lebrun, et al. 2001) until telomere shortening during mating-type switching inactivates them (fig. 1 below). This permits conversion of the *MATa* gene, which is situated closer to the centromere, to the *MATα* gene in the subtelomeric region (Klar 2007). Although mice cells embody a master-and proto-silencer mechanism, silencers are interspersed throughout the HoxD cluster and involved in the timing of HoxD gene expression (Fourel, et al. 2002). Significantly, not only do the yeast *MATa* and mouse HoxD occupy similar chromosomal positions relative to the centromere, but the homeodomains of encoded proteins are identical in three-dimensional structure (Gehring 1998). Thus, *MATa* and *MATα* may represent "proto"-Hox orthologues. From the foregoing we predict 1) that continued analysis of metazoan subtelomeric regions will demonstrate that they mirror *S. cerevisiae* in having master-and proto-silencers whose unmasking via telomere shortening allows them to act on their counterparts in associated Hox clusters and 2), since the significantly longer telomeres of higher eukaryotes shorten in "steps," the subtelomeric master-and proto-silencers will be unmasked in a sequential, stepwise fashion, resulting in an ordered expression of Hox genes (fig. 1 below). Further, given that ARS consensus sequence-containing proto-silencers convert to anti silencers in replication-factor *S. cerevisiae* mutants (Rehman, et al. 2009), developmental anomalies reported for humans can be understood in terms of nucleotide alteration of subtelomeric master-and/or proto-silencers leading to abnormal Hox gene expression.

**Telomeres and metazoan form** Our hypothesis – asymmetrical shortening of telomeres due to the down regulation of telomerase represents an external counting mechanism that enables cellular differentiation and thus complex form – gains support from the following:

When the system reactivates telomerase, telomere shortening and the process of chromosomal-length asymmetry cease, and cells become cancerous, wherein descendent lineages do not maintain a differentiated state (Artandi and DePinho 2010). Since, like prokaryotic and normally telomerase-positive lower eukaryotic cell lineages, metazoan telomerase-positive, cancer-cell lineages are essentially "immortal," it is obvious that cells of a lineage cannot simultaneously be telomerase positive and negative (allowing differentiation) (Di Giacomo, et al. 2013). Further, although coupled with telomere length and regulated by shared proteins (e.g. TRF1 in mammals), cell-cycle control is independent of Hox gene expression. When TRF1 is not bound to telomeres, as in late G2 and M mitosis, it interacts with cell-cycle proteins to regulate entry into the M phase (Zhou, et al. 2003). Additionally, targeted deletion of Trf1 in mice leads to early embryonic lethality (Karlseder, et al. 2003). Consequently, telomere shortening can be conceived as an independent and irreversible counting mechanism that is external not only to Hox genes, but also to GRN systems. In turn, metazoans must express a homogeneous and highly organized developmental program that is coupled with Hox gene expression and implemented by GRNs. Studies on null mice lacking telomerase activity in the TERT (Wong et al., 2003) or TERC (Blasco, et al. 1997) components also support our hypothesis. These mice suffer developmental defects in multiple tissues. In successive crossbred generations, normally highly proliferative tissues become increasingly compromised; by the sixth generation, offspring are infertile (Herrera, et al. 1999). This suggests that while the counting mechanism is partially defective in early generations, it can still enable cellular differentiation, albeit with consequent developmental abnormalities. By the sixth generation, telomeres are too short to "count," and development is derailed. As summarized above, we propose that, if a mutational event results in telomere extension coincident followed by cells becoming telomerase negative, subsequent cell divisions will give rise to one cell lineage through which an unaltered copy of the genome will be retained (Di Giacomo, et al. 2013) (fig. 1 above). In a telomerase down-regulated "organism," successive cell division cycles together with duplication and mutation of chromosomal genes coincident with the emergence of primitive Hox genes would yield a genome capable of a hierarchy of differential expression. And it is such a novel genomic configuration that may have contributed to the origin of complex, metazoan form (Di Giacomo, et al. 2013).

**Conclusion: further speculation** Building upon McAdams and Shapiro's (McAdams and Shapiro 1995) suggestion that, during development, GRNs are equivalent to the Boolean Logic Circuits (BLCs) on which computer hard-and soft-ware are based, Davidson (Davidson 2010) argued that GRNs can be conceived as consecutive BLCs that emerge over time via transformation of preceding circuits. These characterizations are, however, inaccurate. A BLC cannot transform itself into a new circuit or acquire a new gate or function (Stankovic and Astola 2011). Moreover, even if GRNs were controlled by the successive activation of Hox genes, and GRNs and Hox gene expression were mutually interactive, the shift from one GRN to another or one Hox gene to another cannot be internally orchestrated. It must be coupled and synchronized with an *external* counting mechanism because, while a computer program may be a nexus of BLCs, something external to that program must jumpstart it (= bootstrapping). Thus, if one invokes a computer analogy, one cannot co-opt only part of it. One must consider the entire architecture and its embedded mathematical logic. Recently, Brenner (Brenner 2012) suggested correspondence between the (universal) Turing machine model and a cell. However, since a single-celled organism or a terminally differen-tiated cell merely copies itself (i.e. expresses the same genes), rather than expressing recursive functions or referring from within to an external anchor, it cannot represent a universal Turing machine. If, however, the process of metazoan development embodies both an external reference point (i.e. a counting mechanism) and the property not of iteration, but of recursivity (i.e. cells creating different versions of themselves by using different gene sets from the same genome), it could represent a universal Turing machine suitable for modeling the development of any metazoan. If recursivity is not a property of metazoan development, it may still be possible to demonstrate that this process reflects a hierarchy of special Turing machines. In either case, the potential of our hypothesis lies in identifying the irreversible, sequential expression of sets of genes during metazoan development as a reduction of information that is progressively converted via increasing cellular differentiation into complex organismal form: i.e. the emergence of metazoan complexity derives from trading information for structure (Csuhaj-Varjú, et al. 2008). Thus, in contrast to Brenner's (Brenner 2012) claim that biology is in crisis, we suggest that integrative perspectives – perhaps as presented here –are moving the field into a new and exciting phase of inquiry.

**Figure legend Fig. 1**. **Above**: Asymmetrical shortening of a telomeric end through three cell divisions; L: average length (AvL) of the telomere considering lagging and leading strands; X: extent of telomere shortening per cell division due to discontinuous synthesis of the lagging strand in one of two daughter cells. Colors identify telomeres by average length. Below: Similarity between *S. cerevisiae* mating-type switching and mouse HoxD de-silencing. **A.** In *S. cerevisiae*, the HMLα silent-mating cassette is silenced by the telomere and protosilencers (TPE). With telomere shortening, HMLα is de-silenced and gene conver-

sion takes place with the MATa gene. **B.** In mice embryos, temporal de-silencing of HoxD genes is a function of telomere shortening (arrows show transcription of HoxD1 and HoxD3 genes). Since more than one cell division may be necessary to de-silence a single HoxD gene, we present a simplified diagram: Subtelomeric repetitive sequences are temporally de-silenced by telomere shortening and act on master and protosilencer counterparts interspersed in the HoxD cluster. The Hox gene cluster is equivalent to a sequential access memory that codes for the main instructions of the developmental program, while all other chromosomal regions (not represented) are equivalent to a random access memory that codes for housekeeping genes and GRNs. This organization resembles the Harvard computer architecture.

**References**

Artandi SE, DePinho RA 2010. Telomeres and telomerase in cancer. Carcinogensis 31: 9-18.

Blasco MA, Lee HW, Hande MP, Samer E, Lansdorp PM, DePinho RA, Greider CW 1997. Telomere shortening and tumor formation by mouse cells lacking telomerase RNA. Cell 91: 25-34.

Brenner S 2012. Turing centenary: life's code script. Nature 482: 461.

Csuhaj-Varjú E, Di Nola A, Păun G, Pérez-Jiménez MJ, Vassal G 2008. Editing configurations of P systems. Fundamenta Informaticae 82: 29-46.

Davidson EH 2010. Emerging properties of animal gene regulatory networks. Nature 468: 911-920.

Davidson EH, Erwin DH 2006. Gene regulatory networks and the evolution of animal body plans. Science 311: 796-800.

Di Giacomo R, Schwartz JH, Maresca B 2013. The origin of Metazoa: an algorithmic view of 224 life. Biological Theory 8: 221-231.

Duboule D 1994. Temporal colinearity and the phylotypic progression: a basis for the stability of a vertebrate Bauplan and the evolution of morphologies through heterochrony. Development (Supplement) 1994: 135-142.

Fourel G, Lebrun E, Gilson E 2002. Protosilencers as building blocks for heterochromatin. 229 BioEssays 24: 828-835.

Gehring W. 1998. Master control genes in development and evolution: the homeobox story. 231 Yale University Press: New Haven.

Herrera E, Samper E, Blasco MA 1999. Telomere shortening in mTRK/K embryoes is associated with failure to close the neural tube. EMBO 18: 1172-1181.

Heutink P, et al. 1994. The gene for triphalangeal thumb maps to the subtelomeric region of chromosome 7q. Nature Genetics 6: 287-292.

Huffman KE, Leven SD, Tesmer VM, Shay JW, Wright WE 2000. Telomer shortening is proportional to the size of the G-rich telomeric 3'-overhang. Journal of Biological Chemistry 275: 19719-19722.

Imai S, Armstrong CM, Kaeberlein M, Guarente L 2000. Transcriptional silencing and longevity protein Sir2 is an NAD-dependent histone deacetylase. Nature 403: 795-800.

Karlseder J, Kachatrian L, Takai H, Mercer K, Hingorani S, Jacks T, de Lange T 2003. Targeted deletion reveals an essential function for the telomere length regulator Trf1. Molecular and Cell Biology 23: 6533-6541.

Klar AJS 2007. Lessons learned from studies of fission yeast mating-type switching and silencing. Annual Review of Genetics 41: 213-236.

Lebrun E, Revardel E, Boscheron C, Li R, Gilson E, Fourel G 2001. Protosilencers in *Saccharomyces cerevisiae* subtelomeric regions. Genetics 158: 167-176.

Mallo M, Wellik DM, Deschamps J 2010. Hox genes and ergional patterning of the vertebrate body plan. Developmental Biology 344: 7-15.

McAdams HH, Shapiro L 1995. Circuit simulation of genetic networks. Science 269: 650-656.

Michishita E, McCord RA, Boxer LD, Barber MF, Hong T, Gozani O, Chua KF 2009. Cell cycle-dependent deacetylation of telomeric histone H3 lysine K56 by human SIRT6. Cell 253 Cycle 8: 2664-2669.

O'Hare TH, Delany ME 2005. Telomerase gene expression in the chicken: telomerase RNA (TR) and reverse trasncriptase (TERT) transcript profiles are tissue-specific and correlate with telomerase activity. Age 27: 631-641.

Okabe-Oho Y, Murakami H, Oho S, Sasai M. 2009. Stable, precise, and reproducible patterning of bicoid and hunchback molecules in the early *Drosophila* embryo. PLoS Computational Biology. doi: e1000486.

Quinonez SC, Innis JW 2014. Human *HOX* gene disorder. Molecular Genetics and Metabolism 111: 4-15. 262 Rehman M, A,, Wang D, Fourel G, Gilson E, Yankulov K 2009. Subtelomeric ACS-containing 263 Proto-silencers Act as Antisilencers in Replication Factors Mutants in *Saccharomyces* 264 *cerevisiae*. Molecular Biology of the Cell 20: 631-641.

Schultz C, Tautz D 1995. Zygotic caudal regulation by hunback and its role in abdominal segment formation of the *Drosophila* embryo. Development 121: 1023-1028.

Stankovic RS, Astola J. 2011. From Boolean Logic to Switching Circuits and Automata: towards modern information technology. Springer-Verlag: Berlin.

Taylor HA, Delany ME 2000. Ontogeny of telomerase in chicken: impact of down-regulation on pre-and postnatal telomer length in vivo. Development. Growth and Differentiation 42: 271 613-621.

Tschopp P, Duboule D 2011. A genetic approach to the transcriptional regulation of Hox gene clusters. Annual Review of Genetics 45: 145-166.

Wartlick O, Kicheva A, González-Gaitán M 2009. Gradient formation. Cold Spring Harbor Perspectives on Biology 1. doi: a001255

Wolpert L 2010. Arms and the man: the problem of symmetric growth. PLoS Biology 8: 1-3.

Wolpert L 1969. Positional information and the spatial pattern of cellular differentiation. Journal of Theoretical Biology 25: 1-45.

Wolpert L 2011. Positional informaton and patterning revisited. Journal of Theoretical Biology 269: 359-365.

Wolpert L 2007. Specifying positional information in the embryo: looking beyond morphogens. Cell 130: 205-209.

Yan D, Lin X 2009. Shaping morphogen gradients by proteoglycans. Cold Spring Harbor Perspectives on Biology 1. doi: (3)a002493

Zhou XZ, Perrem K, Ping Lu K 2003. Role of Pin2/TRF1 in telomere maintenance and cell cycle control. Journal of Cell Biochemstriy 89: 19-37. 287 288

**Figure 1**

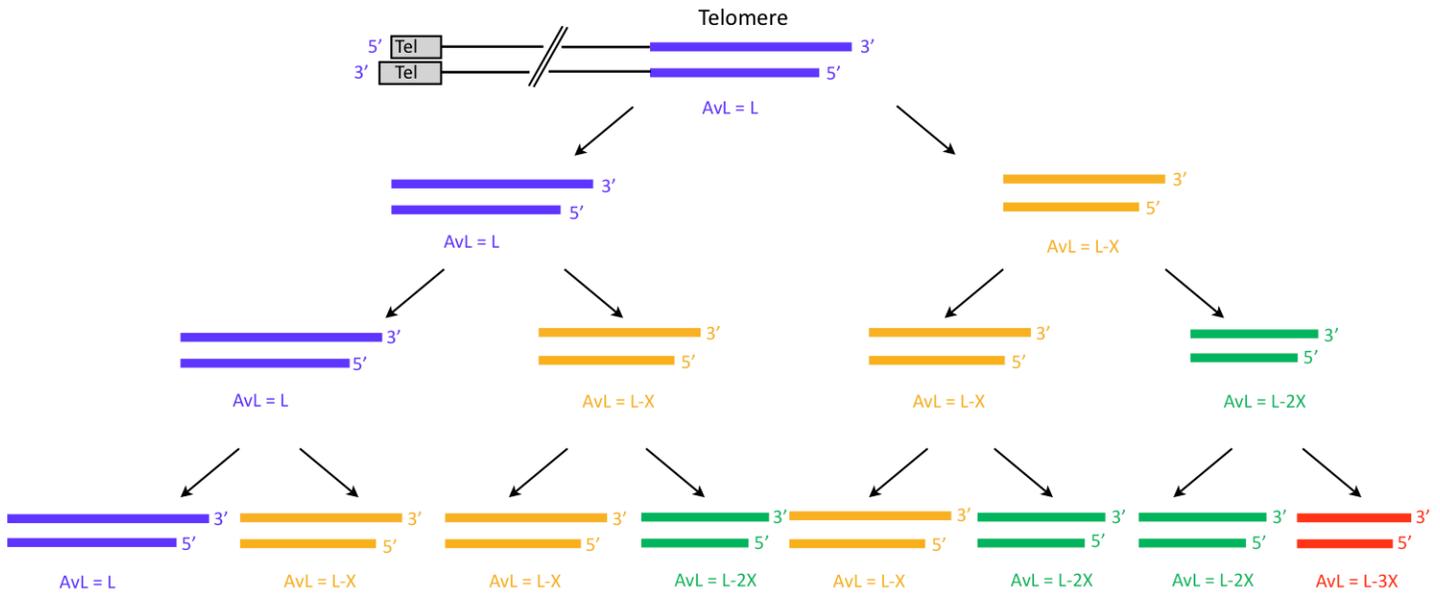

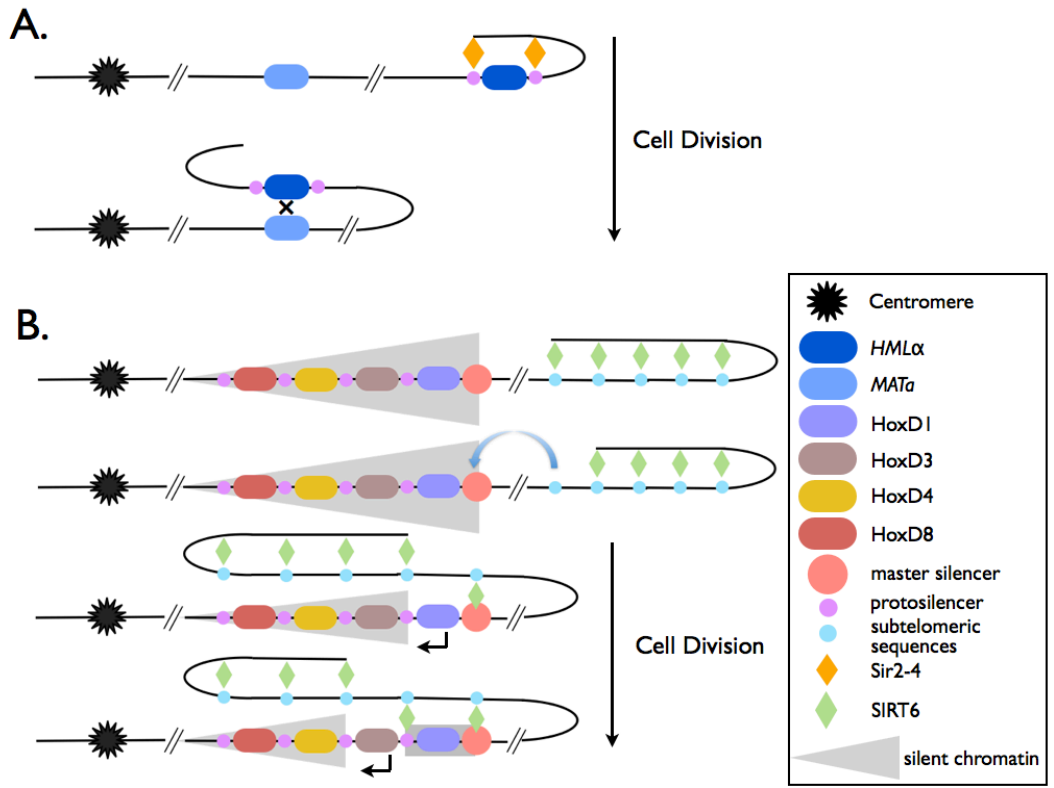